\documentclass[letterpaper, 10 pt, conference]{ieeeconf}  

\IEEEoverridecommandlockouts                              

\usepackage{amsmath} 
\usepackage{amssymb}  
\usepackage[utf8]{inputenc}
\usepackage[T1]{fontenc}
\usepackage{microtype}
\usepackage[dvips]{graphicx}
\usepackage{xcolor}
\usepackage{times}
\usepackage{float}
\usepackage{mathabx}
\usepackage{dsfont}
\usepackage[ruled,vlined,noresetcount]{algorithm2e}
\usepackage{cancel}
\usepackage{multirow}
\usepackage{blindtext}
\usepackage{amssymb}
\usepackage{tikz}
\usepackage{url}
\usetikzlibrary{shapes,arrows,calc,positioning}

\newcommand\norm[1]{\left\lVert#1\right\rVert}

\title{\LARGE \bf
Control of Mechanical Systems via Feedback Linearization\\Based on Black-Box Gaussian Process Models}

\author{Alberto Dalla Libera$^{1*}$, Fabio Amadio$^{1*}$, Daniel Nikovski$^{2}$, Ruggero Carli$^{1*}$ and Diego Romeres$^{2}$
\thanks{$^{1}$ Alberto Dalla Libera, Fabio Amadio and Ruggero Carli are with department of Information Engeneering, University of Padova, Via Gradenigo 6/B,
35131 Padova, Italy {\tt\small[dallaliber@dei.unipd.it, amadiofa@dei.unipd.it, carlirug@dei.unipd.it]}}%
\thanks{$^{2}$ Diego Romeres and Daniel Nikovski are with Mitsubishi  Electric Research Laboratories (MERL), Cambridge, MA 02139 {\tt\small [romeres@merl.com, nikovski@merl.com]}}%
\thanks{$^*$Alberto Dalla Libera, Fabio Amadio and Ruggero Carli were partially supported by MIUR (Italian Ministry for Education) under the initiative “Departments of Excellence” (Law 232/2016)}
}%

\begin{document}

\maketitle
\thispagestyle{empty}
\pagestyle{empty}

\begin{abstract}

In this paper, we consider the use of black-box Gaussian process (GP)  models for trajectory tracking control based on feedback linearization, in the context of mechanical systems. We considered two strategies. The first computes the control input directly by using the GP model, whereas the second computes the input after estimating the individual components of the dynamics. We tested the two strategies on a simulated manipulator with seven degrees of freedom, also varying the GP kernel choice. Results show that the second implementation is more robust w.r.t. the kernel choice and model inaccuracies. Moreover, as regards the choice of kernel, the obtained performance shows that the use of a structured kernel, such as a polynomial kernel, is advantageous, because of its effectiveness with both strategies.

\end{abstract}

\section{INTRODUCTION}
Dynamics models are fundamental in robotics. For instance, inverse dynamics models, which relate joint trajectories to joint torques, are used in high-precision trajectory tracking applications \cite{FL_control,FL_control_elastic,siciliano}, and also in problems where robots interact with the environment, such as force control \cite{force_control,siciliano}, impedance control \cite{impedance_control,impedance_control_non_diag_stiff}, and collision detection \cite{col_det,col_det_GP}. 

In the aforementioned applications, the accuracy of the inverse dynamics model is crucial. However, deriving an accurate model of the robot inverse dynamics is a challenging task, in particular when system specifications are limited or uncertain, or when complex behaviors such as friction and elasticity are relevant. Indeed, in these contexts, the  identification of parametric models derived from first principles of physics \cite{parametric_ID,handbook} are often not effective, due to model bias and unmodeled behaviors. For these reasons, in the last decades, several black-box and grey-box strategies for inverse dynamics identification have been proposed. A relevant class of solutions is based on Gaussian Process Regression (GPR) \cite{rasmussen_GP_for_ML}, see for instance \cite{GIP,SP_peters,romeres2016online,Sparse_GP_RT_control, Rezaei_cascaded_GP}. Here, instead of identifying the physical parameters of the model, the inverse dynamics are treated as an unknown function, which relates position, velocity, and acceleration of the joints to torques. This unknown function is modeled a priori as a Gaussian Process (GP), with covariance parametrized through a kernel function \cite{rasmussen_GP_for_ML, Scholkopf_LWK}. The posterior distribution of the joint torques, given the observed data, can be computed in closed form, and can be used to predict joint torques.

\begin{figure}
	\centering
	\includegraphics[width=0.9\linewidth , keepaspectratio]{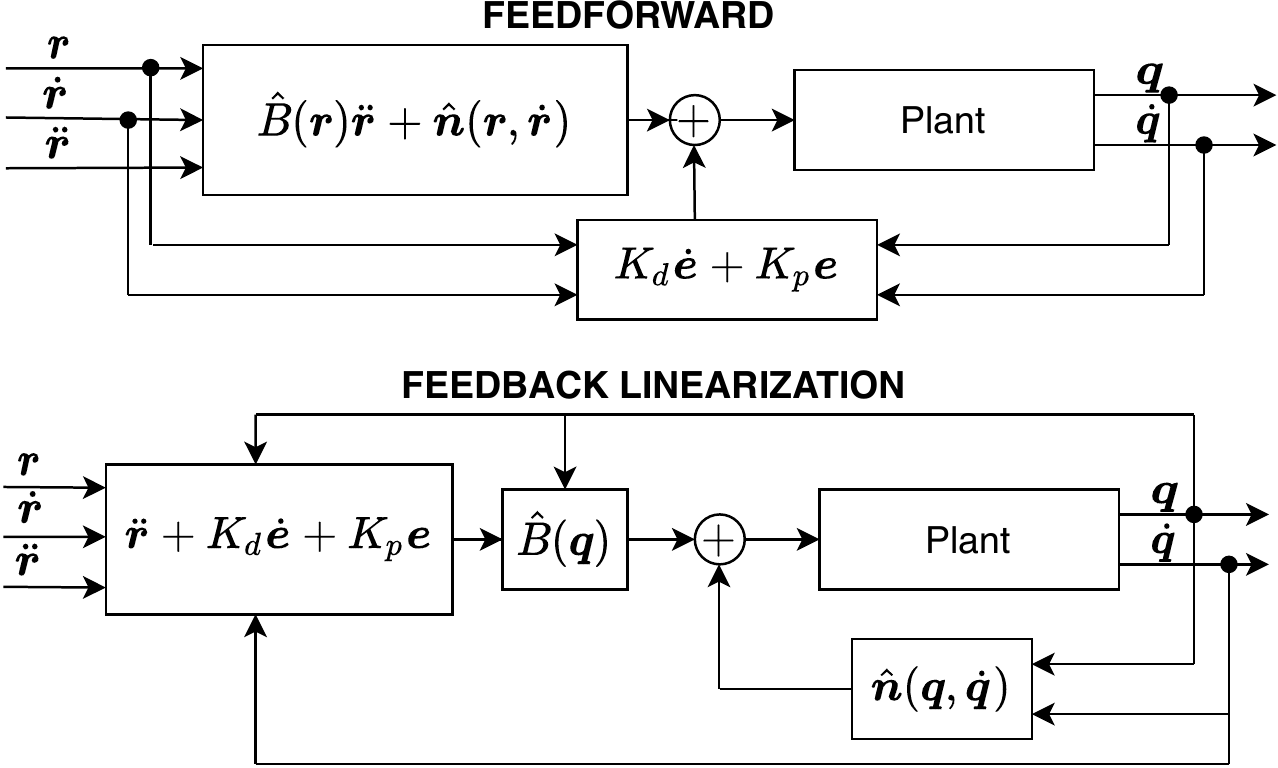}
	\caption{Diagrams of feedforward and feedback linearization control. $\boldsymbol{q}$, $\boldsymbol{r}$, and $\boldsymbol{e}$ are the joints position, the reference trajectory, and the tracking error.}
	\label{fig:control_diagrams}
\end{figure}

Compared to physical models, which are strictly related to the dynamics equations, GP models are less interpretable, and, consequently, their use in control applications might be less straightforward. However, several works show that such models can be used in applications, see, for instance, \cite{Sparse_GP_RT_control,Peters_computed_torque_control,stulp_computed_torque_control,hirche_stable_GP_control_4_lagrangian, Rezaei_cascaded_GP} concerning trajectory tracking, and \cite{col_det_GP} concerning proprioceptive collision detection. Typically, in trajectory tracking, GP models are exploited by implementing a feedforward control scheme \cite{craig_intro_robotics}, see the diagram in Figure \ref{fig:control_diagrams}. Instead of using parametric models, in the GP implementation, the feedforward term is the output of the GP model evaluated for the position, velocity, and acceleration of the reference trajectory. The control loop is closed with a decentralized PD controller to cancel errors. When the GP model is accurate and the PD gains are set properly, the feedback loop is effective in canceling the residual tracking error. However, there are some issues that could limit the performance of a feedforward controller, as follows. (i) In the feedback loop, coupling between different degrees of freedom (DoF) are not considered. (ii) The robot inertia is configuration dependent, and, in some cases, it might be difficult to obtain a set of PD gains that can assure the same performance for all configurations. (iii) Convergence is not guaranteed, even if the inverse dynamics are known exactly; see \cite{craig_intro_robotics}, chapter 10.5 for details about (ii) and (iii).

An alternative control scheme is feedback linearization control \cite{FL_control,FL_control_elastic,siciliano}, described in the diagram in Figure \ref{fig:control_diagrams}. In contrast to feedforward control, where the model is used to compute a proper control input in advance, in feedback linearization, the inverse dynamics model is used to obtain a tracking error with linear dynamics. The control input is the sum of two terms. The first aims at compensating all the torques independent of accelerations. The second is given by a feedforward term proportional to the reference acceleration, and a PD feedback term. To account for couplings and variations of the inertia matrix, the second term is computed using an estimate of the inertia matrix. In contrast to feedforward control, feedback linearization assures asymptotic convergence, if the dynamics are known exactly. Moreover, the error dynamics are described by a second-order linear differential equation, fully characterized by the PD gains, providing a principled way to set the PD gains \cite{siciliano}. 

In this work, we analyze two implementations of feedback linearization control based on GP models. The first implementation is simpler, and estimates directly the feedback linearization control input using the GP model. In contrast, the second implementation is composed of two steps. First, the inertia matrix and the compensation of the torques independent of accelerations are estimated separately by means of the GP model. Then, the feedback linearization control input is computed by applying its standard form. To the best of our knowledge, the first implementation has been attempted before only in \cite{Peters_computed_torque_control}. However, that paper was focused on issues related to modeling, and not to control. In contrast, the second implementation has never been proposed before, and it requires the estimation of several different components of the dynamics equations from the GP model, which is introduced in this paper. We tested the two implementations with a simulated 7-DoF manipulator, varying also the choice of the GP prior, i.e., its kernel. The obtained results show that the second implementation is more robust w.r.t. the kernel choice and initial errors.

The remainder of the paper is organized as follows. In Section \ref{sec:background}, we provide background formulations of robot dynamics and control, as well as GPR. Section \ref{sec:GP_dyn_equations} describes the strategy proposed to estimate several different dynamics components from black-box GP models, and in Section \ref{sec:GP_FL} we describe the two feedback linearization algorithms implemented. Experiments are described in Section \ref{sec:experiments}, and conclusions are drawn in Section \ref{sec:conclusions}.

\section{BACKGROUND} \label{sec:background}
In the first part of this section, we provide background formulation of robot dynamics, as well as introduce the trajectory tracking problem and describe the feedforward and the feedback linearization controllers. In the second part, we describe GPR for inverse dynamics identification, detailing  the black-box priors adopted in this work.

\subsection{Robot dynamics and control}
Consider a mechanical systems with $n$ degrees of freedom, and denote with $\boldsymbol{q}_t \in \mathbb{R}^n$ its generalized coordinates at time $t$; $\dot{\boldsymbol{q}}_t$ and $\ddot{\boldsymbol{q}}_t$ are the velocity and the acceleration of the joints, respectively. The generalized torques, i.e., the control input of the system, are denoted with $\boldsymbol{\tau}_t \in \mathbb{R}^n$. For compactness, in the following, we will denote explicitly the dependencies on $t$ only when strictly necessary. Under rigid body assumptions, the dynamics equations of a mechanical system are described by the following matrix equation
\begin{equation}\label{eq:dyn_eq}
    B(\boldsymbol{q}) \ddot{\boldsymbol{q}} + \boldsymbol{c}(\boldsymbol{q}, \dot{\boldsymbol{q}}) + \boldsymbol{g}(\boldsymbol{q}) + \boldsymbol{F}(\dot{\boldsymbol{q}})  = \boldsymbol{\tau} \text{,}
\end{equation}
where $B(\boldsymbol{q})$ is the inertia matrix, while $\boldsymbol{c}(\boldsymbol{q}, \dot{\boldsymbol{q}})$, $\boldsymbol{g}(\boldsymbol{q})$, and $\boldsymbol{F}(\dot{\boldsymbol{q}})$ account, respectively, for the contributions of fictitious forces, gravity, and friction, see \cite{siciliano} for a more detailed description. For compactness, we introduce also $\boldsymbol{n}(\boldsymbol{q}, \dot{\boldsymbol{q}}) = \boldsymbol{c}(\boldsymbol{q}, \dot{\boldsymbol{q}}) +\boldsymbol{g}(\boldsymbol{q})+\boldsymbol{F}(\dot{\boldsymbol{q}})$. In the following, we will denote with $\hat{B}(\boldsymbol{q})$ and $\hat{\boldsymbol{n}}(\boldsymbol{q}, \dot{\boldsymbol{q}})$ the estimates of $B(\boldsymbol{q})$ and $\boldsymbol{n}(\boldsymbol{q}, \dot{\boldsymbol{q}})$.

The trajectory tracking problem consists in designing a controller able to follow a reference trajectory $\boldsymbol{r}_t,\dot{\boldsymbol{r}}_t,\ddot{\boldsymbol{r}}_t$, starting from initial conditions $\boldsymbol{q}_{t_0},\dot{\boldsymbol{q}}_{t_0},\ddot{\boldsymbol{q}}_{t_0}$. 


In feedback linearization control, the control input $\boldsymbol{\tau}$ is
\begin{subequations}\label{eq:FL_eq}
\begin{align}
    &\boldsymbol{a} = \ddot{\boldsymbol{r}}  + K_p \boldsymbol{e} + K_d \dot{\boldsymbol{e}} \label{eq:FL_acc}\text{,}\\
    &\boldsymbol{\tau} = \hat{B}(\boldsymbol{q}) \boldsymbol{a} + \hat{\boldsymbol{n}}(\boldsymbol{q}, \dot{\boldsymbol{q}}) \label{eq:FL_tau}\text{.}
\end{align}
\end{subequations}
Assuming that the model is known exactly, i.e., $\hat{B}(\boldsymbol{q}) = B(\boldsymbol{q})$ and $\hat{\boldsymbol{n}}(\boldsymbol{q}, \dot{\boldsymbol{q}}) = \boldsymbol{n}(\boldsymbol{q}, \dot{\boldsymbol{q}})$, combining \eqref{eq:dyn_eq} and \eqref{eq:FL_eq} and recalling that $B(\boldsymbol{q})$ is invertible, it can be proven that the tracking error goes asymptotically to zero if $K_p>0$ and $K_d>0$ \cite{siciliano}. Indeed, under these assumptions, the dynamics of the tracking error is described by the following second order linear differential equation
\begin{equation}\label{eq:FL_error_dyn}
\ddot{\boldsymbol{e}} + K_d \dot{\boldsymbol{e}} + K_p \boldsymbol{e} = 0 \text{,}
\end{equation}
which is stable if $K_p>0$ and $K_d>0$. This fact represent a considerable advantage w.r.t. feedforward control, since it provides a principled way to chose $K_p$ and $K_d$. Indeed, selecting $K_p = \omega^2 I$ and $K_d=2\zeta\omega I$, with $I$ being the identity matrix, we obtain $n$ decoupled second-order input/output relations with natural frequency $\omega$ and damping ratio  $\zeta$.

\subsection{GPR for inverse dynamics identification}
GPR provides a solid probabilistic framework to identify the inverse dynamics from data. Typically, in GPR, each joint torque is modeled by a distinct and independent GP. Consider an input/output dataset $\mathcal{D} = \left\{\boldsymbol{y}^{(i)}, X \right\}$, where $\boldsymbol{y}^{(i)} \in \mathbb{R}^N$ is a vector collecting $N$ measurements of $\tau^{(i)}$, the \emph{i}-th joint torque, while $X=\left\{\boldsymbol{x}_{t_1}\dots\boldsymbol{x}_{t_N}\right\}$; $\boldsymbol{x}_{t}$ is the vector collecting the position, velocity and acceleration of the joints at time $t$, hereafter denoted GP input. The probabilistic model of $\mathcal{D}$ is
\begin{equation*}
    \boldsymbol{y}^{(i)} =
    \begin{bmatrix}
    f^{(i)}\left(\boldsymbol{x}_{t_1}\right) \\ \vdots \\f^{(i)}\left(\boldsymbol{x}_{t_N}\right)
    \end{bmatrix}
    + \begin{bmatrix}
    w^{(i)}_{t_1} \\ \vdots \\ w^{(i)}_{t_N}
    \end{bmatrix}
      = \boldsymbol{f}^{(i)}(X) + \boldsymbol{w}^{(i)} \text{,}
\end{equation*}
where $\boldsymbol{w}^{(i)}$ is i.i.d. Gaussian noise with standard deviation $\sigma_i$, while $f^{(i)}(\cdot)$ is an unknown function modeled a priori as a GP, namely, $f^{(i)}(\cdot) \sim N(0,\mathbb{K}^{(i)}(X,X))$. The covariance matrix $\mathbb{K}^{(i)}(X,X)$ is defined through a kernel function $k^{(i)}(\cdot, \cdot)$. Specifically, the covariance between $f^{(i)}\left(\boldsymbol{x}_{t_j}\right)$ and $f^{(i)}\left(\boldsymbol{x}_{t_l}\right)$, i.e., the element of $\mathbb{K}^{(i)}(X,X)$ at row \emph{j} and column \emph{l}, is equal to $k^{(i)}\big(\boldsymbol{x}_{t_j}, \boldsymbol{x}_{t_l}\big)$. Exploiting the properties of Gaussian distributions, it can be proven that the posterior distribution of $f^{(i)}$ given $\mathcal{D}$ in a general input location $\boldsymbol{x}_{*}$ is Gaussian \cite{rasmussen_GP_for_ML}. Then, the maximum a posteriori estimator corresponds to the mean, which is given by the following expression
\begin{equation}\label{eq:GP_estimate}
    \hat{f}^{(i)}(\boldsymbol{x}_*) = \mathbb{K}^{(i)}\left(\boldsymbol{x}_*,X\right)\boldsymbol{\alpha}^{(i)} \text{,}
\end{equation}
where
\begin{align*}
    &\boldsymbol{\alpha}^{(i)} = (\mathbb{K}^{(i)}\left(X,X\right) + \sigma_i^2 I)^{-1}\boldsymbol{y}^{(i)} \text{,}\\
    &\mathbb{K}^{(i)}\big(\boldsymbol{x}_*,X\big) = \left[k^{(i)}\big(\boldsymbol{x}_*, \boldsymbol{x}_{t_1}\big) \dots k^{(i)}\big(\boldsymbol{x}_*, \boldsymbol{x}_{t_N}\big)\right] \text{.}
\end{align*}

Different solutions proposed in the literature can be grouped roughly based on the definition of the GP prior. In this paper, we will consider two black-box approaches, where the prior is defined without exploiting prior information about the physical model.

\textbf{Squared Exponential kernel} The Squared Exponential (SE) kernel \cite{rasmussen_GP_for_ML,Scholkopf_LWK}, defines the covariance between samples based on the distance between GP inputs, and it is defined by the following expression
\begin{equation}\label{eq:SE_kernel}
    k_{SE}\big(\boldsymbol{x}_{t_j}, \boldsymbol{x}_{t_l}\big) = \lambda e^{-\norm{\boldsymbol{x}_{t_j}-\boldsymbol{x}_{t_l}}^{2}_{\Sigma}} \text{;}
\end{equation}
$\lambda$ and $\Sigma$ are the kernel hyperparameters. The first is a scaling factor, and the second is a positive definite matrix, which defines the norm used to compute the distance between inputs. A common choice consists in considering $\Sigma$ to be diagonal, with the positive diagonal elements named lengthscales.

\textbf{Geometrically Inspired Polynomial kernel} The Geometrically Inspired Polynomial (GIP) kernel has been recently introduced in \cite{GIP}. This kernel is based on the property that the dynamics equations in \eqref{eq:dyn_eq} are a polynomial function in a proper transformation of the GP input, fully characterized only by the type of each joint. Specifically, $\boldsymbol{q}$ is mapped in $\tilde{\boldsymbol{q}}$, the vector composed by the concatenation of the components associated with a prismatic joint and the sines and cosines of the revolute coordinates. As proved in \cite{GIP}, the inverse dynamics in \eqref{eq:dyn_eq} is a polynomial function in $\ddot{\boldsymbol{q}}$, $\dot{\boldsymbol{q}}$ and $\tilde{\boldsymbol{q}}$, where the elements of  $\ddot{\boldsymbol{q}}$ have maximum relative degree of one, whereas the ones of $\dot{\boldsymbol{q}}$ and $\tilde{\boldsymbol{q}}$ have maximum relative degree two. To exploit this property, the GIP kernel is defined through the sum and the product of different polynomial kernels \cite{MPK}, hereafter denoted as $k_P^{(p)}(\cdot,\cdot)$, where $p$ is the degree of the polynomial kernel. In particular, we have 
\begin{align}
    &k_{GIP}\big(\boldsymbol{x}_{t_j}, \boldsymbol{x}_{t_l}\big) = \label{eq:GIP_eq}\\ 
    &\left(k_P^{(1)}\big(\ddot{\boldsymbol{q}}_{t_j}, \ddot{\boldsymbol{q}}_{t_l}\big) + k_P^{(2)}\big(\dot{\boldsymbol{q}}_{t_j}, \dot{\boldsymbol{q}}_{t_l}\big)\right)
    k_Q\big(\tilde{\boldsymbol{q}}_{t_j}, \tilde{\boldsymbol{q}}_{t_l}\big) \text{,} \nonumber
\end{align}
where, in its turn, $k_Q$ is given by the product of polynomial kernels with degree two, see \cite{GIP} for all the details. In this way, the GIP kernel allows defining a regression problem in a finite-dimensional function space where \eqref{eq:dyn_eq} is contained, leading to better data efficiency in comparison with the SE kernel.

\section{ESTIMATE OF THE DYNAMICS COMPONENTS FROM GAUSSIAN PROCESS MODELS OF THE INVERSE DYNAMICS} \label{sec:GP_dyn_equations}
In this section, we describe how it is possible to obtain estimates of the different contributions in the left-hand side of \eqref{eq:dyn_eq} when adopting GPR to identify the inverse dynamics; in particular, we discuss the computation of gravitational contributions, inertial contributions, and $\boldsymbol{n}(\boldsymbol{q},\dot{\boldsymbol{q}})$. We assume that a distinct GP is used for each of the $n$ degree of freedom, and we denote by $\hat{f}^{(i)}(\cdot)$, $i=1\dots n$, the estimator of the \emph{i}-th joint torque obtained applying \eqref{eq:GP_estimate}. For convenience, from here on, we will point out explicitly the different components of the GP input, namely, the input of the $\hat{f}^{(i)}$ will be $(\boldsymbol{q},\boldsymbol{\dot{q}},\boldsymbol{\ddot{q}})$ instead of $\boldsymbol{x}$, which comprises the concatenation of $\boldsymbol{q},\boldsymbol{\dot{q}},\boldsymbol{\ddot{q}}$. It is worth mentioning that the proposed approach is inspired by the strategy adopted in Newton-Euler algorithms, see \cite{deluca_NE_dyn_est}.

\subsection{Gravitational contribution}
As shown in \eqref{eq:dyn_eq}, the torques due to the gravitational contributions account for all the terms that depend only on $\boldsymbol{q}$. Consequently, to obtain $g^{(i)}(\boldsymbol{q})$, i.e., the estimate of the \emph{i}-th gravitational contribution in the configuration $\boldsymbol{q}$, we evaluate $\hat{f}^{(i)}$ by setting $\boldsymbol{\dot{q}}=\boldsymbol{0}$, $\boldsymbol{\ddot{q}}=\boldsymbol{0}$. Then, the estimate of $\boldsymbol{g}(\boldsymbol{q})$ is 
\begin{equation}\label{eq:g_estimate}
    \hat{\boldsymbol{g}}(\boldsymbol{q}) =
    \begin{bmatrix}
    \hat{g}^{(1)}(\boldsymbol{q})\\
    \vdots\\
    \hat{g}^{(n)}(\boldsymbol{q})
    \end{bmatrix}
    =
    \begin{bmatrix}
    \hat{f}^{(1)}(\boldsymbol{q},\boldsymbol{0},\boldsymbol{0})\\
    \vdots\\
    \hat{f}^{(n)}(\boldsymbol{q},\boldsymbol{0},\boldsymbol{0})
    \end{bmatrix} \text{.}
\end{equation}

\subsection{Inertial contributions}
The inertial contributions, i.e., $B(\boldsymbol{q})\boldsymbol{\ddot{q}}$, accounts for all the contributions that depend simultaneously on $\boldsymbol{q}$ and $\boldsymbol{\ddot{q}}$. Consequently, to estimates these contributions, we evaluate the GP models in $(\boldsymbol{\ddot{q}}, \boldsymbol{0}, \boldsymbol{q})$, and subtract the gravitational contribution defined and computed previously. In particular, to obtain $\hat{B}_{ij}(\boldsymbol{q})$, i.e., the estimate of the $B(\boldsymbol{q})$ element in position $(i,j)$, we set all the accelerations to zero, except for the \emph{j}-th component. Denoting with $\boldsymbol{1}_j$ the vector with all elements equal to zero except for the $j$-th element, which, instead, is equal to one, we have
\begin{equation}\label{eq:B_ij_estimate}
   \hat{B}_{ij}(\boldsymbol{q}) = \hat{f}^{(i)}(\boldsymbol{q},\boldsymbol{0},\boldsymbol{1}_j) - \hat{g}^{(i)}(\boldsymbol{q}) \text{.}
\end{equation}

\subsection{Estimation of $\boldsymbol{n}(\boldsymbol{q})$}
The vector $\boldsymbol{n}(\boldsymbol{q})$ collects all the contributions that do not depend on $\ddot{\boldsymbol{q}}$. Then, $n^{(i)}(\boldsymbol{q},\dot{\boldsymbol{q}})$, i.e., the estimate of the \emph{i}-th component of $\boldsymbol{n}(\boldsymbol{q})$, is computed by evaluating the \emph{i}-th GP model setting $\ddot{\boldsymbol{q}}=\boldsymbol{0}$. Then, we have
\begin{equation}\label{eq:n_estimate}
    \hat{\boldsymbol{n}}(\boldsymbol{q}, \dot{\boldsymbol{q}}) =
    \begin{bmatrix}
    \hat{n}^{(1)}(\boldsymbol{q}, \dot{\boldsymbol{q}})\\
    \vdots\\
    \hat{n}^{(n)}((\boldsymbol{q}, \dot{\boldsymbol{q}}))
    \end{bmatrix}
    =
    \begin{bmatrix}
    \hat{f}^{(1)}(\boldsymbol{q},\dot{\boldsymbol{q}},\boldsymbol{0})\\
    \vdots\\
    \hat{f}^{(n)}(\boldsymbol{q},\dot{\boldsymbol{q}},\boldsymbol{0})
    \end{bmatrix} \text{.}
\end{equation}

\section{FEEDBACK LINEARIZATION CONTROL BASED ON GAUSSIAN PROCESS MODEL}\label{sec:GP_FL}
In this section, we describe the two GP-based feedback linearization controllers implemented. The first implementation aims at estimating directly an approximation of \eqref{eq:FL_eq} using the GP models, whereas the second computes the approximation of \eqref{eq:FL_eq} by estimating $B(\boldsymbol{q})$ and $\boldsymbol{n}(\boldsymbol{q},\dot{\boldsymbol{q}})$ using the expressions derived in Section \ref{sec:GP_dyn_equations}. 

\subsection{GP-FL}\label{sec:GP_FL_NO_B}
In this approach, hereafter denoted as GP Feedback Linearization control (GP-FL), the control input is selected to be directly the estimate of \eqref{eq:FL_tau}. The estimate of \eqref{eq:FL_tau} at time $t$ is obtained by evaluating the $n$ GP models with GP-input given by the concatenation of $\boldsymbol{q}_t$, $\dot{\boldsymbol{q}}_t$ and $\boldsymbol{a}_t = \ddot{\boldsymbol{r}}_t + K_p \boldsymbol{e}_t +K_d \dot{\boldsymbol{e}}_t$. Then, referring to the notation previously introduced, we have
\begin{equation}\label{eq:GPFL}
    \boldsymbol{\tau}_t =
    \begin{bmatrix}
    \hat{f}^{(1)}(\boldsymbol{q}_t,\dot{\boldsymbol{q}}_t,\boldsymbol{a}_t)&
    \dots&
    \hat{f}^{(n)}(\boldsymbol{q}_t,\dot{\boldsymbol{q}}_t,\boldsymbol{a}_t)
    \end{bmatrix}^T \text{.}
\end{equation}

\subsection{GP-FL-DCE}\label{sec:GP_FL_with_inertia}
The second approach, named GP Feedback Linearization control with Dynamics Components Estimation (GP-FL-DCE), computes the control input based on \eqref{eq:FL_eq} and the estimation of $B(\boldsymbol{q})$ and $\boldsymbol{n}(\boldsymbol{q},\dot{\boldsymbol{q}})$ obtained with the GP input. First, the elements of the inertia matrix and the estimates of $\boldsymbol{n}(\boldsymbol{q},\dot{\boldsymbol{q}})$ are computed by applying, respectively, \eqref{eq:B_ij_estimate} and \eqref{eq:n_estimate}. Then, the input is
\begin{equation}\label{eq:GPFLDCE}
    \hat{\boldsymbol{\tau}}_t=
    \begin{bmatrix}
    \hat{B}_{11}(\boldsymbol{q}) & \dots & \hat{B}_{1n}(\boldsymbol{q})\\
    \vdots & \vdots & \vdots\\
    \hat{B}_{n1}(\boldsymbol{q}) & \dots & \hat{B}_{nn}(\boldsymbol{q})
    \end{bmatrix}
    \boldsymbol{a}_t + 
    \begin{bmatrix}
    \hat{n}^{(1)}(\boldsymbol{q}, \dot{\boldsymbol{q}})\\
    \vdots\\
    \hat{n}^{(n)}((\boldsymbol{q}, \dot{\boldsymbol{q}}))
    \end{bmatrix}\text{,}
\end{equation}
where, as before, $\boldsymbol{a}_t = \ddot{\boldsymbol{r}}_t + K_p \boldsymbol{e}_t +K_d \dot{\boldsymbol{e}}_t$.

\section{EXPERIMENTS}\label{sec:experiments}
Experiments have been carried out in PyBullet \cite{pybullet}, simulating a KUKA LBR iiwa, which is a 7-DoF collaborative manipulator\footnote{A video of the experiments is available at \url{https://youtu.be/ehy8iDRGIDo}}. The system was controlled at 1000 HZ. Position, velocity, and torques of the joints are directly provided by the simulator. The accelerations needed for model identification were computed offline by means of acausal numerical differentiation of the velocities. Specifically, we applied the central difference approximation, namely, the acceleration of the joints at time $t$ is approximated with $\ddot{\boldsymbol{q}}_t = (\dot{\boldsymbol{q}}_{t+1}-\dot{\boldsymbol{q}}_{t-1})/(2\delta T)$, where $\delta T$ is the sampling time.

The remainder of this section is organized as follows. First, we compare the accuracy of GP models obtained with the SE and GIP kernel. Second, the control performance of GP-FL and GP-FL-DCE is compared on a trajectory tracking problem, with initial tracking error equal to zero and varying the kernel choice. Finally, the two strategies are tested on the same trajectory tracking problem, in the presence of initial tracking errors for all joints.

\subsection{Model learning performance}\label{exp_accuracy}
To train and test the GP models obtained with the SE and GIP kernel, we collected two data sets, hereafter denoted by $\mathcal{D}_{tr}$ and $\mathcal{D}_{tests}$. The first data set, $\mathcal{D}_{tr}$, was used to derive the GP estimators \eqref{eq:GP_estimate}, after optimizing the kernel hyperparameters by marginal likelihood maximization \cite{rasmussen_GP_for_ML}. The second data set, $\mathcal{D}_{test}$, was used to compare the performance of the two GP estimators. Both data sets were collected by employing a hand-tuned PD controller to track a random reference trajectory. For each joint, the reference trajectory was Gaussian noise filtered with a second-order low-pass filter (with cutoff frequency 1 Hz). The length of the trajectory was 50 seconds, resulting in 50,000 samples. To limit the computational complexity of \eqref{eq:GP_estimate}, the collected samples were down-sampled with a constant rate of 10, obtaining 5,000 samples for each dataset. 

\begin{figure}
	\centering
	\includegraphics[width=0.9\linewidth , keepaspectratio]{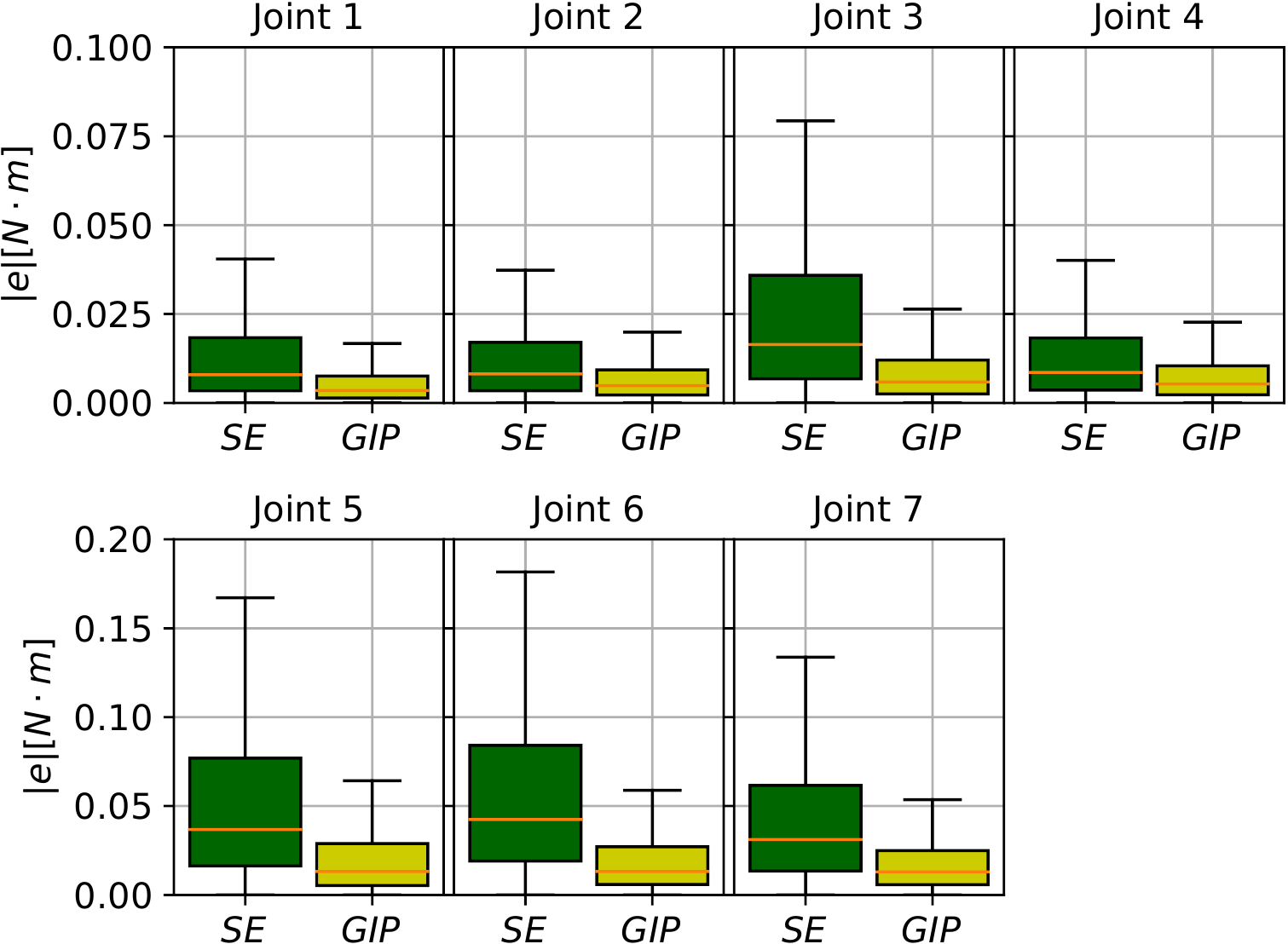}
	\caption{Boxplots of the absolute errors obtained in $\mathcal{D}_{test}$ with the SE and GIP kernels. In the table below, we report the nMSE percentages.}
	\smallskip
	\begin{small}
	\begin{tabular}{|l|l|l|l|l|l|l|l|}
  \hline
  kernel & $\tau_1$ & $\tau_2$ & $\tau_3$ & $\tau_4$ & $\tau_5$ & $\tau_6$ & $\tau_7$\\
  \hline
  SE & 3.99 & 0.48 &  4.22 &  0.88 &  7.95 & 10.86 &
  4.91\\
  \hline
  GIP & 0.42 & 0.12 & 0.55 & 0.18 & 1.21 & 1.80 &
 1.05\\
  \hline
\end{tabular}
\end{small}
	\label{fig:estimate_perf}
\end{figure}

In Figure \ref{fig:estimate_perf}, we visualize the distribution of the absolute value of the errors  obtained in $\mathcal{D}_{test}$ with the two GP estimators. Moreover, in the table below Figure   \ref{fig:estimate_perf}, we reported the normalized Mean Squared Error (nMSE), namely, the ratio between the mean squared error and the variance of the correspondent joint torques, expressed as a percentage. As already showed in \cite{GIP}, for all joints, the estimator based on the GIP kernel outperforms the one based on the SE kernel, showing better data efficiency and generalization. 

\subsection{Trajectory tracking without initial tracking error}\label{sec:exp_no_error}
\begin{figure*}[!h]
	\centering
	\includegraphics[width=0.9\textwidth , keepaspectratio]{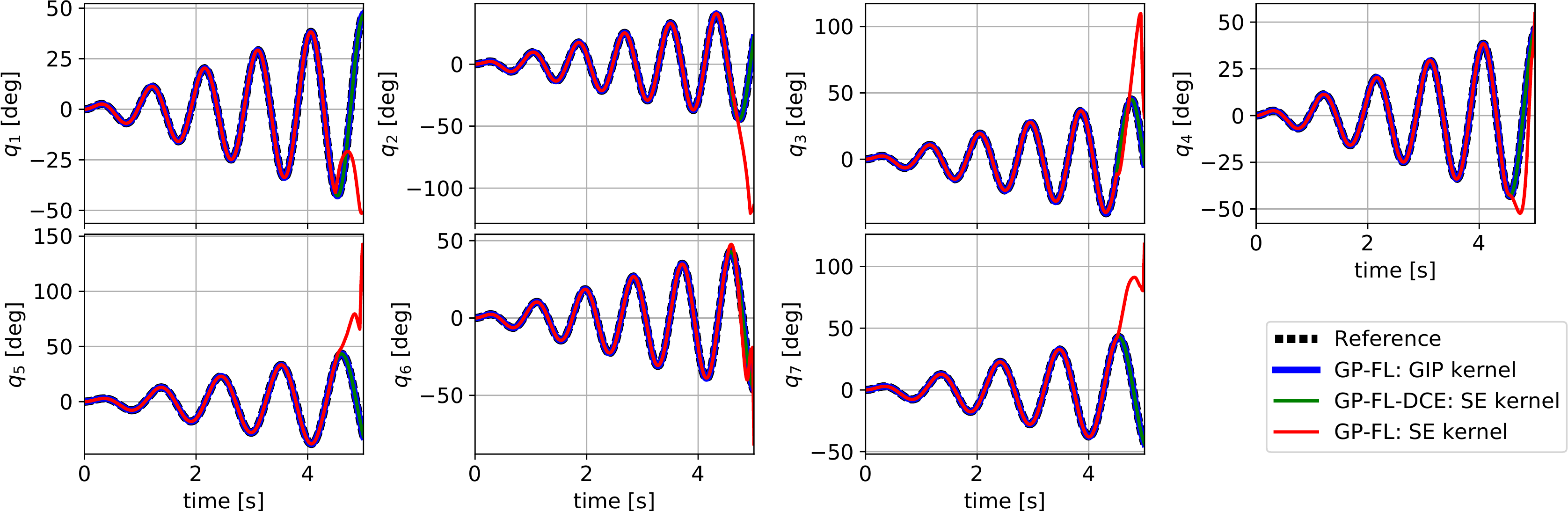}
	\caption{Joints trajectories obtained with GP-FL (with SE and GIP kernel), and GP-FL-DCE (with the SE kernel) in experiment of Section \ref{sec:exp_no_error}.}
	\label{fig: tracking}
\end{figure*}
\begin{figure*}[h]
	\centering
	\includegraphics[width=0.9\textwidth , keepaspectratio]{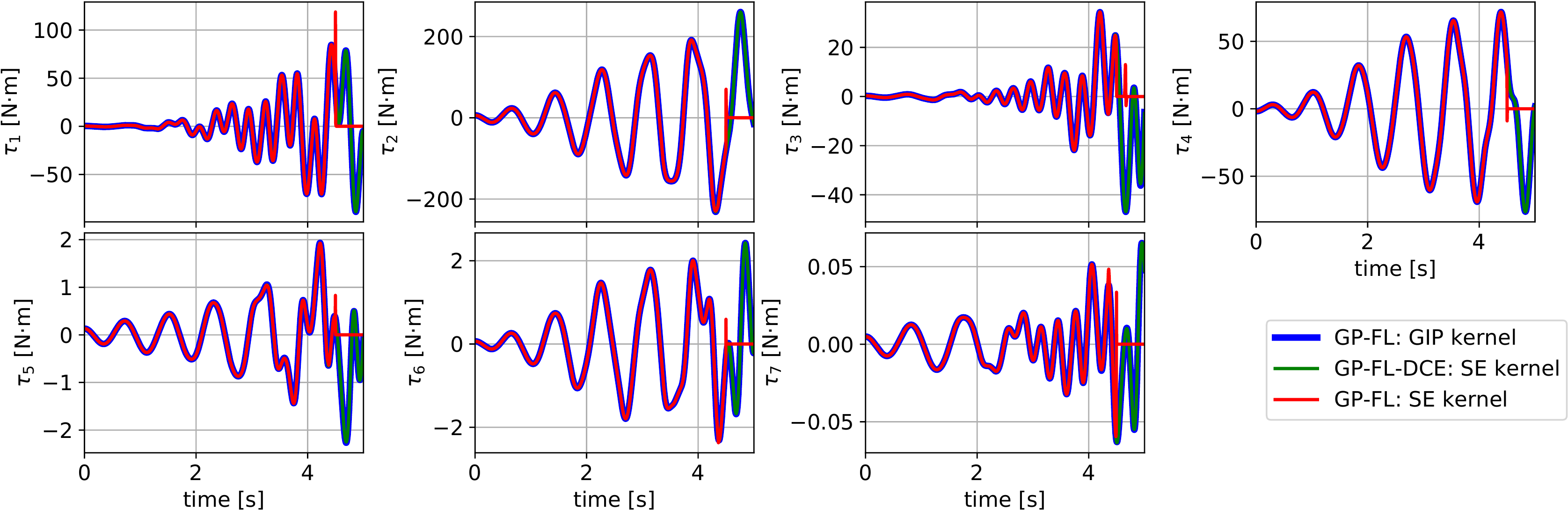}
	\caption{Joint torques obtained with GP-FL (with SE and GIP kernel), and GP-FL-DCE (with the SE kernel) in experiment of Section \ref{sec:exp_no_error}.}
	\label{fig: torques}
\end{figure*}
In the first control experiment, the GP-FL and GP-FL-DCE controllers based on the two models were tested on the same trajectory tracking problem. For each dof, $j=1,\dots,7$, the reference joint position was given by $r_t^{(j)}=0.165\, t\, sin(2\pi F_j t)$, where the frequencies $F_j$ were randomly sampled from $\mathcal{N}(0.5, 1)$. The controller gains are selected following the  considerations reported in Section \ref{sec:background}, $K_p = \omega^2I$ and $K_d=2\zeta\omega I$, with $\omega=100$ and $\zeta=2$. The control horizon was 5s, and the initial tracking error was zero. In Figure \ref{fig: tracking} and \ref{fig: torques}, the evolution of the joint angles and control torques obtained by GP-FL with SE and GIP kernel, and GP-FL-DCE with SE kernel are reported, respectively. 

First, we discuss the performance obtained using the model based on the SE kernel. It can be noticed that the GP-FL controller with the SE kernel works properly when the amplitudes of the reference oscillations are low, but it starts to fail suddenly towards the end of the control horizon, when zero torques are commanded to all joints. This observation suggests that the GP-FL scheme evaluates the GP model in unexplored regions, where predictions are equal to the prior mean \cite{rasmussen_GP_for_ML}, which is zero. This is due to the large magnitude of $\boldsymbol{a}_t$, which grows with the tracking error, and becomes significantly different from the accelerations seen during training. Instead, GP-FL-DCE with the SE kernel is able to track the reference trajectory, demonstrating better robustness compared to the GP-FL. This robustness is likely achieved thanks to the estimation of the individual components of the dynamics. Indeed, even though the robot is far from the configurations seen during training, the GP model based on SE provides sufficiently accurate estimates of $B(\boldsymbol{q})$ and $\boldsymbol{n}(\boldsymbol{q})$, which results in keeping the robot close to the reference. 

Thanks to the better generalization of the GIP kernel, the GP-FL controller based on the GIP kernel is  more robust and is always able to track the desired reference trajectory, also in unexplored areas of the state space. The performance of GP-FL-DCE with the GIP kernel has not been reported, since the trajectory obtained with this scheme and GP-FL-DCE with the GIP kernel is the same of GP-FL with GIP kernel. This is due to the definition of the GIP kernel, which is closer to the physics of the system, and already encodes the linear dependencies of the torques on the acceleration of the joints.

\subsection{Trajectory tracking with initial tracking error}\label{sec:exp_with_error}
\begin{figure*}[!h]
	\centering
	\includegraphics[width=0.9\textwidth , keepaspectratio]{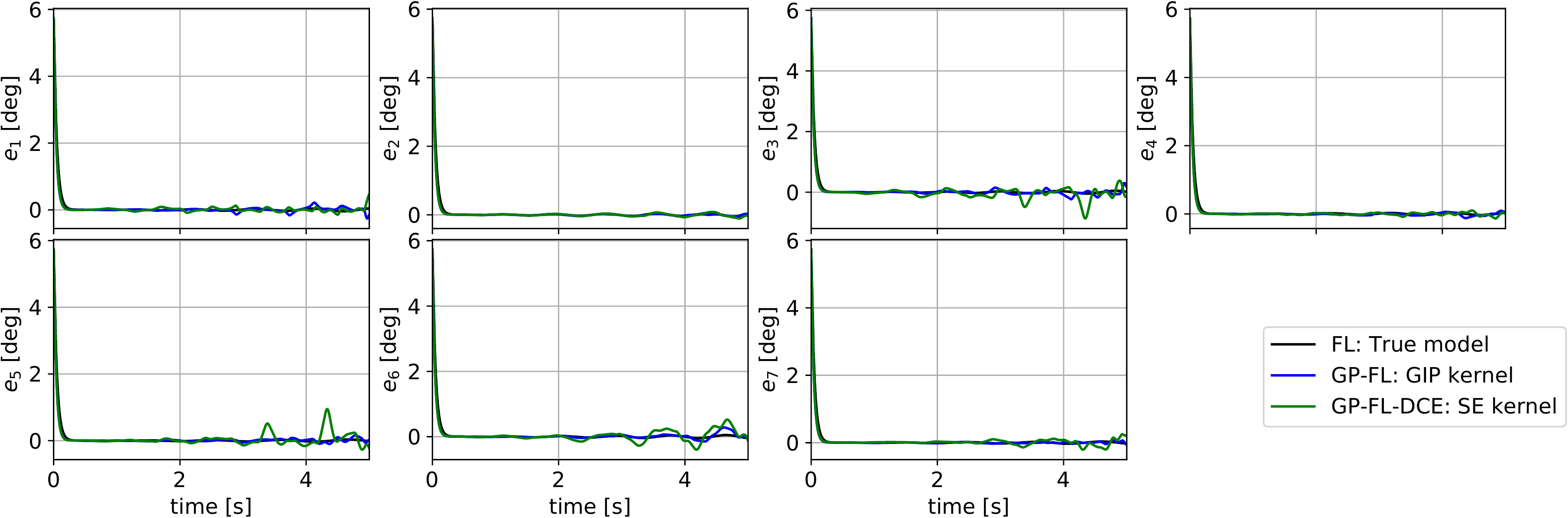}
	\caption{Tracking errors obtained with feedback linearization based on the true model, GP-FL (with GIP kernel), and GP-FL-DCE (with SE kernel).}
	\label{fig: errors}
\end{figure*}
In this experiment, we tested the controllers on the same reference trajectory as in the previous experiment, considering also the presence of initial tracking errors. For all joints, we considered an initial error of $5.73^o$. The obtained behavior confirmed the observations from the previous experiment. The GP-FL scheme with the SE kernel is not effective. In fact, the initial error makes the magnitude of the $\boldsymbol{a}_t$ term large, leading to considerable distances from accelerations observed during training, and zero torques from the beginning. In Figure \ref{fig: errors}, we plotted the tracking errors obtained by GP-FL-DCE with the SE kernel and GP-FL with the GIP kernel, as well as the one obtained by a feedback linearization control based on the true model. For all three estimators, the main dynamics of the tracking error follows the exponential behavior described in \eqref{eq:FL_error_dyn}. Significant differences between the tracking error evolution can be appreciated only at steady state, where the controllers based on GP models are subject to limited oscillations around zero, with absolute value lower than $1^o$, and growing with the amplitude of the reference trajectories. These errors are due to model inaccuracies, which becomes more relevant when the reference trajectories cross regions that are far from the distribution of the training samples. The errors are higher for the controller based on the SE kernel. This is in accordance with the considerations presented in Section \ref{exp_accuracy}, where we highlight that the model based on GIP is more accurate. In particular, the tracking errors at steady state are higher in joint 3, 4, 6, and 7, which are the ones where the GP estimator is less accurate, as confirmed by the nMSE obtained in the experiment of Section \ref{exp_accuracy}.

\section{CONCLUSIONS}\label{sec:conclusions}
In this paper, we analyze the implementation of feedback linearization control based on GP models. We considered two strategies. The first computes the control input directly with the GP model, whereas the second computes the input after estimating the individual components of the dynamics, in particular, the inertia matrix and the torques independent of accelerations. The two strategies were compared on a trajectory tracking problem with a simulated 7-DoF manipulator, varying also the kernel choice; we considered the SE and GIP kernels. Results show that the second implementation is more robust w.r.t. the kernel choice and model inaccuracies. Moreover, as regards the choice of kernel, the obtained performance shows that the use of a structure kernel, such as the GIP kernel, is advantageous, resulting in good performance for both implementations.

\bibliographystyle{IEEEtran}
\bibliography{references}

\end{document}